\documentclass[a4paper,twoside,12pt]{article}
\input{obsjkt.sty}

\newcommand{\Msun}{\ensuremath{~{\rm M}_\odot}}                   
\newcommand{\Rsun}{\ensuremath{~{\rm R}_\odot}}                   
\newcommand{\rhosun}{\ensuremath{~\rho_\odot}}                    
\newcommand{\Teff}{\ensuremath{T_{\rm eff}}}                      
\newcommand{\Porb}{\ensuremath{P_{\rm orb}}}                      
\newcommand{\EBV}{\ensuremath{E(B\!-\!V)}}                        
\newcommand{\Grp}{\ensuremath{G_{\rm RP}}}                        
\newcommand{\degr}{\ensuremath{^\circ}}                           
\renewcommand{\kms}{~km~s$^{-1}$}                                 
\renewcommand{\cd}{~d$^{-1}$}                                     
\newcommand{\chir}{\ensuremath{\chi_\nu^{\,2}}}                   
\newcommand{\etal}{\textit{et al.}}                               
\newcommand{\gaia}{\textit{Gaia}}                                 
\newcommand{\targ}{AR~Aur}
\newcommand{\targfull}{AR~Aurigae}

\newcommand{\Msunnom}{\hbox{$\mathcal{M}^{\rm N}_\odot$}}
\newcommand{\Rsunnom}{\hbox{$\mathcal{R}^{\rm N}_\odot$}}
\newcommand{\Lsunnom}{\hbox{$\mathcal{L}^{\rm N}_\odot$}}

\usepackage{rotating}

\begin{document} 

\OBSheader{Rediscussion of eclipsing binaries: \targ}{J.\ Southworth}{2025 June}

\OBStitle{Rediscussion of eclipsing binaries. Paper XXV. \\ The chemically-peculiar system AR~Aurigae}

\OBSauth{John Southworth}

\OBSinstone{Astrophysics Group, Keele University, Staffordshire, ST5 5BG, UK}


\OBSabstract{\targ\ is a detached eclipsing binary containing two late-B stars which are chemically peculiar, on a circular orbit of period 4.135~d. The primary is a HgMn star which shows temporal changes in its chemical abundances and spectral line profiles, whilst the secondary is a likely weak Am star. Published analyses of the system have used spectroscopic light ratios to constrain the eclipse models and found that the secondary star is larger than the primary. This unexpected outcome has been taken as an indication that the system is young and the secondary has yet to reach the main sequence. In this work we present the first analysis of the light curve of the system obtained by the Transiting Exoplanet Survey Satellite (TESS), whose quality allows us to avoid using a spectroscopic light ratio to constrain the solution. When combined with literature spectroscopic results we obtain highly precise masses of $2.544 \pm 0.009$\Msun\ and $2.358 \pm 0.009$\Msun\ and radii of $1.843 \pm 0.002$\Rsun\ and $1.766 \pm 0.003$\Rsun. The light ratio is inconsistent with spectroscopic determinations, confirming the suggestion of Takeda \cite{Takeda25raa} that spectroscopic light ratios of the system are unreliable due the chemical peculiarity of the stars. The properties of the system are matched by theoretical predictions for a slightly super-solar metallicity and an age of $33 \pm 3$~Myr: both components are young main-sequence stars.}


\section*{Introduction}

The detached eclipsing binary system (dEB) \targfull\ has been suggested to be a young object in which the secondary component is still a pre-main-sequence star \cite{NordstromJohansen94aa2}. This claim was based on the less massive star having a larger radius and lower surface gravity, caused by using a spectroscopic light ratio (SLR) as a constraint in the eclipse modelling. A recent work by Takeda \cite{Takeda25raa} questioned this claim because at least one of the stars is chemically peculiar, making light ratios from spectroscopic absorption lines unreliable. In this work we present an analysis of a new space-based light curve which does not use an SLR as a constraint, confirms the suggestion by Takeda, and yields improved measurements of the physical properties of the \targ\ system.

The current work is presented in the context of our series of papers which revisit known dEBs \cite{Me20obs} for which higher-quality light curves are now available \cite{Me21univ}. The ultimate aim is to measure the masses and radii of the component stars to 2\% precision \cite{Andersen91aarv,Torres++10aarv} and enable their inclusion in the Detached Eclipsing Binary Catalogue \cite{Me15aspc} (DEBCat\footnote{\texttt{https://www.astro.keele.ac.uk/jkt/debcat/}}).


\section*{\targfull}

\begin{table}[t]
\caption{\em Basic information on \targfull. 
The $BV$ magnitudes are each the mean of 111 individual measurements \cite{Hog+00aa} distributed approximately randomly in orbital phase. 
The $JHK_s$ magnitudes are from 2MASS \cite{Cutri+03book} and were obtained at an orbital phase of 0.23. \label{tab:info}}
\centering
\begin{tabular}{lll}
{\em Property}                            & {\em Value}                 & {\em Reference}                      \\[3pt]
Right ascension (J2000)                   & 05 18 18.896                & \citenum{Gaia23aa}                   \\
Declination (J2000)                       & +33 46 02.52                & \citenum{Gaia23aa}                   \\
Bright Star Catalogue                     & HR 1728                     & \citenum{HoffleitJaschek91}          \\
Henry Draper designation                  & HD 34364                    & \citenum{CannonPickering18anhar2}    \\
\textit{Hipparcos} designation            & HIP 24740                   & \citenum{Hipparcos97}                \\
\textit{Tycho} designation                & TYC 2398-1311-1             & \citenum{Hog+00aa}                   \\
\textit{Gaia} DR3 designation             & 181983575426242944          & \citenum{Gaia21aa}                   \\
\textit{Gaia} DR3 parallax (mas)          & $7.0735 \pm 0.0461$         & \citenum{Gaia21aa}                   \\          
TESS\ Input Catalog designation           & TIC 144085463               & \citenum{Stassun+19aj}               \\
$B$ magnitude                             & $6.102 \pm 0.014$           & \citenum{Hog+00aa}                   \\          
$V$ magnitude                             & $6.144 \pm 0.010$           & \citenum{Hog+00aa}                   \\          
$J$ magnitude                             & $6.190 \pm 0.019$           & \citenum{Cutri+03book}               \\
$H$ magnitude                             & $6.254 \pm 0.017$           & \citenum{Cutri+03book}               \\
$K_s$ magnitude                           & $6.265 \pm 0.023$           & \citenum{Cutri+03book}               \\
Spectral type                             & B9 V + B9.5 V               & \citenum{NordstromJohansen94aa2}     \\[3pt]
\end{tabular}
\end{table}

\targ\ (Table~\ref{tab:info}) has a long observational history, and was the first dEB known in which one component is a chemically-peculiar star of the HgMn type. The discovery of eclipses was made in 1931 by Pedersen \& Steengaard \cite{PedersenSteengaard31bz}, who subsequently measured an orbital period of $\Porb = 2.076$~d \cite{PedersenSteengaard31bz2}. It was named \targfull\ in Prager's Katalog of 1936. Spectroscopic observations by Harper \cite{Harper+35jrasc} and Wyse \cite{Wyse36pasp} showed that the \Porb\ is double this, provided the first measurements of the velocity amplitudes of the two stars ($K_{\rm A}$ and $K_{\rm B}$), and yielded an SLR of approximately 0.9 from the 4481\,\AA\ and 4549\,\AA\ spectral lines. Nassau \cite{Nassau36aj} confirmed that the primary and secondary eclipses have a slightly different depth, and obtained $\Porb = 4.134581$~d.

Photoelectric photometric studies were made by Huffer \& Eggen \cite{HufferEggen47apj}, who adopted an SLR of $0.86 \pm 0.04$ in their analysis, and Johansen \cite{Johansen70aa} using filters similar to the Str\"omgren $uvby$ system. The data from these two papers were modelled by Cester et al.\ \cite{Cester+78aas2} and similar results obtained. O'Connell \cite{Oconnell79ra} presented $UBV$ photometry and found a change in \Porb. Adelman \cite{Adelman98aas} obtained $uvby$ photometry mostly outside eclipse and found no additional variability.

Nordstr\"om \& Johansen \cite{NordstromJohansen94aa2} (hereafter NJ94) presented a detailed analysis of \targ\ using a precise SLR and the {\sc ebop} code to model the light curves from O'Connell \cite{Oconnell79ra} and Johansen \cite{Johansen70aa}. Radial velocity (RV) measurements were taken from Harper \cite{Harper+35jrasc} and Wyse \cite{Wyse36pasp}. The SLR was obtained by Dr.\ Graham Hill using the Mg\,\textsc{ii} 4481\,\AA\ lines and based on seven spectra taken around quadrature phases. When corrected for the slow change of line strength with effective temperature (\Teff), an SLR of $0.866 \pm 0.018$ was found which gave the ratio of the radii to be $k = R_{\rm B}/R_{\rm A} = 1.020 \pm 0.015$. One outcome of their analysis was that the surface gravity of the secondary star (star~B) was lower than that of the primary (star~A); this was (with caveats) interpreted as indicating the system was young and star~B was still in the final stages of contracting onto the zero-age main sequence.

\subsection*{Spectral characteristics}

The chemical peculiarity of \targ\ was first shown by Wolff \& Wolff \cite{WolffWolff76conf} on the basis of an enhanced Hg\,{\sc ii} 3984\,\AA\ line in star~A. More detailed analysis by Wolff \& Preston \cite{WolffWolff76conf} confirmed that star~A is a HgMn star and found that star~B did not show spectral peculiarities. Takeda \etal\ \cite{Takeda++79pasj} found changes in the strength and profile of the 3984\,\AA\ line of star~A and noted that star~B appeared normal in their spectra. However, Stickland \& Weatherby \cite{SticklandWeatherby84aas} found enhanced Hg\,{\sc ii} in \emph{both} components. Khokhlova \etal\ \cite{Khokhlova+95astl} described star~A as a typical HgMn star and found that star~B showed a different type of chemical peculiarity. Zverko \etal\ \cite{Zverko++97coska} found Mn, Ba and Pt to be overabundant in both stars.

Hubrig et al.\ \cite{Hubrig+06mn} found line-profile variability for many chemical elements in star~A, but none in star~B. The projected rotational velocities for both components were measured as $V \sin i = 22 \pm 1$\kms. In a subsequent analysis, Hubrig et al.\ \cite{Hubrig+10mn} found both stars to have a weak magnetic field from spectropolarimetric observations. These authors also used Doppler tomography to detect strong enhancements of Fe and Y in spots on the surface of star~A. The presence of magnetic fields in HgMn stars has been controversial but several detections now exist \cite{Hubrig+12aa,Hubrig+20mn}.

Folsom et al.\ \cite{Folsom+10mn} presented an extensive analysis of atmospheric properties of \targ. They (re)confirmed the HgMn nature of star~A and that star~B shows weak features of being an Am star. They measured the \Teff\ values of the stars, an SLR consistent with that from NJ94, and precise $K_{\rm A}$ and $K_{\rm B}$ values.

Similar conclusions were obtained by Takeda et al.\ \cite{Takeda+19mn}. The detailed abundance measurements in this and papers mentioned above typically disagree by more than their uncertainties, suggesting that the measured abundances are variable over time. Takeda \cite{Takeda25raa} presented further abundance measurements, obtained precise \Teff\ values, and pointed out that the SLRs found in previous work may be unreliable as both stars are chemically peculiar; it was this work that prompted the current analysis.

\subsection*{Nearby stars}

The multiplicity of \targ\ is of interest. Firstly, it is a member of the Auriga OB1 association \cite{Hoffleit64book}. Secondly, it forms a common-proper-motion pair with the A0p star HR~1732 (IQ~Aur). This was originally found by W.\ P.\ Bidelman, reported by Hoffleit \cite{Hoffleit64book} in the Third Revised Edition of her \textit{Catalogue of Bright Stars}, and confirmed by Sargent \& Eggen \cite{SargentEggen65pasp}. Thirdly, there is a third component on a wider orbit in the system which manifests as changes in the observed \Porb\ of the inner binary.

Guarnieri et al.\ \cite{Guarnieri++75aas} found \Porb\ to be variable from an $O-C$ (observed minus calculated) diagram which showed a parabolic trend of the residuals of a linear fit to the times of mid-eclipse. Zverko et al.\ \cite{Zverko+81ibvs} suggested this was due to the light-time travel effect caused by a third star in a wider orbit. Chochol et al.\ \cite{Chochol+88baicz} found the period of this third body, $P_3$, to be between 24.75 and 27.09~yr. NJ94 fitted the times of minimum light to obtain $P_3 = 24.18 \pm 0.21$~yr, with an amplitude of 0.0094~d and a probable small eccentricity of $e_3 = 0.17$. Albayrak et al.\ \cite{Albayrak++03an} and Zasche \cite{Zasche05apss} have progressively refined the orbital properties of the third body.

Wilson \& Van Hamme \cite{WilsonVanhamme14mn} presented a detailed reanalysis of the \targ\ system. Aside from measuring masses to (a questionable) 0.2\% and radii to 0.5\%, they obtained $P_3 = 23.452 \pm 0.096$~yr and $e_3 = 0.262 \pm 0.023$. They also found the minimum mass of the third body to be $0.5122 \pm 0.0087$\Msun\ -- a single main-sequence star of this mass would be much fainter than either of the eclipsing stars, and if it were a binary or a white dwarf it would be fainter still.


\section*{Photometric observations}


\targ\ was observed in eight sectors (19, 43, 44, 45, 59, 71, 73 and 86) by the NASA Transiting Exoplanet Survey Satellite \cite{Ricker+15jatis} (TESS). In all cases data are available at 120~s cadence and were used for our analysis below. Lower-cadence observations (200, 600 and/or 1800~s) are also available for all sectors but were not used due to their lower time resolution. The data were downloaded from the NASA Mikulski Archive for Space Telescopes (MAST\footnote{\texttt{https://mast.stsci.edu/portal/Mashup/Clients/Mast/Portal.html}}) using the {\sc lightkurve} package \cite{Lightkurve18}. 

\begin{figure}[t] \centering \includegraphics[width=\textwidth]{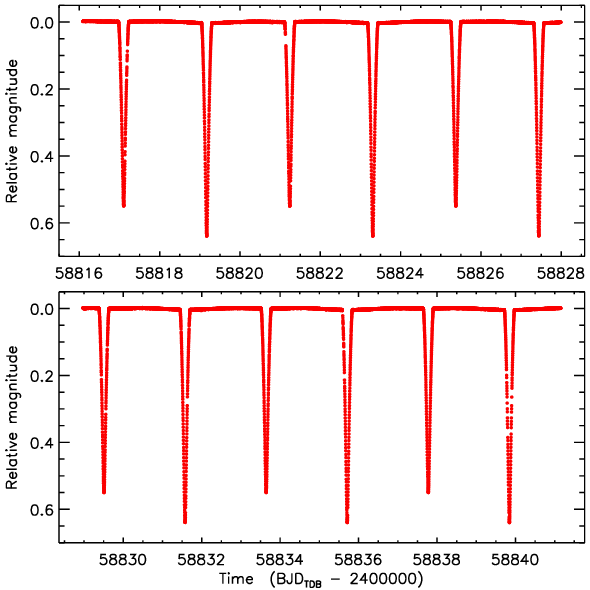} \\
\caption{\label{fig:time} TESS sector 19 photometry of \targ. The flux measurements have 
been converted to magnitude units after which the median was subtracted. The other seven 
sectors used in this work are similar but are not plotted for reasons of space.} \end{figure}

We adopted the simple aperture photometry (SAP) light curves from the SPOC data reduction pipeline \cite{Jenkins+16spie} for our analysis, and rejected low-quality data using the quality flag ``hard''. Additional datapoints from sectors 73 and 86 were rejected manually due to gaps and increased scatter. The remaining data were converted into differential magnitude and the median magnitude was subtracted from each sector for convenience. Fig.~\ref{fig:time} shows the light curve from sector 19; the remaining sectors are similar but for clarity are not plotted.

We queried the \gaia\ DR3 database\footnote{\texttt{https://vizier.cds.unistra.fr/viz-bin/VizieR-3?-source=I/355/gaiadr3}} for all sources within 2~arcmin of \targ. A lot of sources were returned -- 147 -- due to the proximity of the Galactic plane. All are fainter by at least 4.2~mag in the \gaia\ \Grp\ band, so the contamination of the TESS light curve should be small. This is backed up by the \textsc{crowdsap} parameter from TESS, which depends on the sector but is typically in the region of 0.98.


\section*{Light curve analysis}

The components of \targ\ are well-detached and almost spherical, so the light curve is suitable for analysis using the {\sc jktebop}\footnote{\texttt{http://www.astro.keele.ac.uk/jkt/codes/jktebop.html}} code \cite{Me++04mn2,Me13aa}. We modelled the light curves from each sector individually to check for consistency and to guard against small changes in the amount of contaminating light between sectors. We defined star~A to be the star eclipsed at the primary (deeper) eclipse, and star~B to be its companion. These identities are consistent with the literature discussed above.

The fitted parameters were the fractional radii of the stars ($r_{\rm A}$ and $r_{\rm B}$), expressed as their sum ($r_{\rm A}+r_{\rm B}$) and ratio ($k = {r_{\rm B}}/{r_{\rm A}}$), the central surface brightness ratio ($J$), third light ($L_3$), orbital inclination ($i$), orbital period ($P$), and a reference time of primary minimum ($T_0$). A circular orbit was assumed as there is no evidence for orbital eccentricity. Limb darkening (LD) was accounted for using the power-2 law \cite{Hestroffer97aa,Maxted18aa,Me23obs2} and we required both stars to have the same LD coefficients. The linear coefficient ($c$) was fitted and the non-linear coefficient ($\alpha$) fixed at a theoretical value \cite{ClaretSouthworth22aa,ClaretSouthworth23aa}. The observational uncertainties supplied with the TESS flux measurements were scaled to force a reduced $\chi^2$ of $\chir = 1.0$.

\begin{figure}[t] \centering \includegraphics[width=\textwidth]{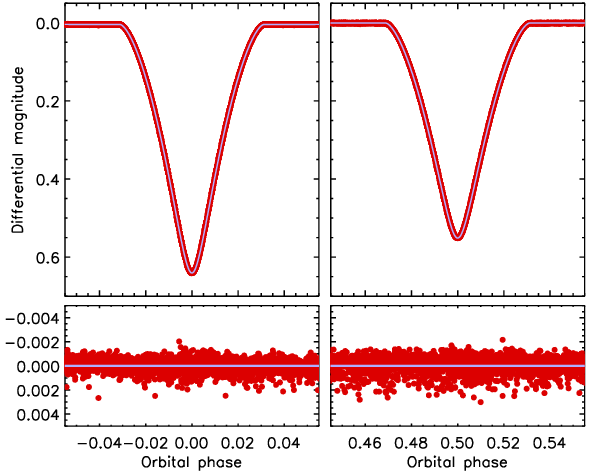} \\
\caption{\label{fig:phase} {\sc jktebop} best fit to the light curves of \targ\ from 
TESS sector 19 for the primary eclipse (left panels) and secondary eclipse (right panels). 
The data are shown as filled red circles and the best fit as a light blue solid line. 
The residuals are shown on an enlarged scale in the lower panels.} \end{figure}

We found that the fits to all sectors were excellent; an example for sector 19 is shown in Fig.~\ref{fig:phase}. The parameters were also highly consistent between sectors, inspiring confidence in the results. In Table~\ref{tab:jktebop} we report the adopted values of the photometric parameters and their uncertainties. We calculated these by taking the unweighted mean and standard deviation of the values for the eight sectors. We did not divide by $\sqrt{8}$ to convert the latter to the standard error as the standard deviations are already very small. We also calculated uncertainties using Monte Carlo and residual-permutation algorithms (tasks 8 and 9 in {\sc jktebop}) and found that their mean values were similar to each other and to the standard deviation.

Our results differ significantly versus previous analyses in that we find star~B to be definitively fainter and smaller than star~A. The radius ratio we find, $0.958 \pm 0.001$, is very different to published values ($1.020 \pm 0.015$ from NJ94 and $1.033 \pm 0.005$ from ref.\ \cite{Folsom+10mn}), and supports the assertion of Takeda \cite{Takeda25raa} that SLRs are not reliable if one or both stars is chemically peculiar. The implications of this result are discussed below.

\begin{figure}[t] \centering \includegraphics[width=\textwidth]{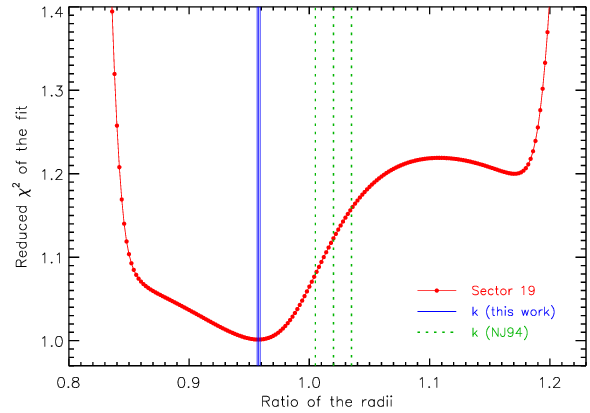} \\
\caption{\label{fig:kfix} Variation of $\chir$ of the {\sc jktebop} fit to the light curve 
from TESS sector 19 as a function of the ratio of the radii $k$ (red line with points). Our 
overall best value and its uncertainty are shown with blue vertical lines, which are very 
close together. The $k$ from NJ94 is shown with vertical green dotted lines.} \end{figure}

To visualise this further we refitted the light curve from sector 19 in the same way as above, but with $k$ fixed at values from 0.8 and 1.2 at intervals of 0.002. The data uncertainties were scaled to give $\chir = 1.0$ for the overall best fit. The result is shown in Fig.~\ref{fig:kfix}, where there is a clear minimum \chir\ corresponding to the adopted value of $k$ in Table~\ref{tab:jktebop}. This $k$ is significantly different to that found by NJ94 and supports our approach of not including an SLR in our light curve fit.

\begin{table} \centering
\caption{\em \label{tab:jktebop} Photometric parameters of \targ\ measured using 
{\sc jktebop} from the light curves from all eight TESS sectors. The errorbars are 
1$\sigma$ and were obtained from the scatter of the results for individual sectors.}
\begin{tabular}{lcc}
{\em Parameter}                           &              {\em Value}            \\[3pt]
{\it Fitted parameters:} \\
Orbital inclination (\degr)               & $      88.6000     \pm  0.0072    $ \\
Sum of the fractional radii               & $       0.19596    \pm  0.00007   $ \\
Ratio of the radii                        & $       0.9578     \pm  0.0013    $ \\
Central surface brightness ratio          & $       0.89939    \pm  0.00032   $ \\
Third light                               & $       0.0152     \pm  0.0027    $ \\
LD coefficient $c$                        & $       0.553      \pm  0.014     $ \\
LD coefficient $\alpha$                   &            0.4318 (fixed)           \\
{\it Derived parameters:} \\
Fractional radius of star~A               & $       0.100089   \pm  0.000052  $ \\       
Fractional radius of star~B               & $       0.095870   \pm  0.000092  $ \\       
Light ratio $\ell_{\rm B}/\ell_{\rm A}$   & $       0.8249     \pm  0.0023    $ \\[3pt]
\end{tabular}
\end{table}

It is beyond the scope of the current work to perform an analysis of the times of minimum light. In Table~\ref{tab:tmin} we report the times of primary mid-eclipse we obtained -- one per sector -- for use by anyone who wishes to do so.

\begin{table} \centering
\caption{\em \label{tab:tmin} Times of minimum light measured for \targ. 
Each time is calculated from the data for a whole sector and corresponds 
to a midpoint of primary eclipse. The final two columns give the uncertainties
calculated via the Monte Carlo and residual-permutation analyses, respectively.}
\begin{tabular}{lccc}
Sector & $T_0$ (BJD$_{\rm TDB}$) & MC error (d) & RP error (d) \\[3pt]
19     & 2458827.440864 & 0.000003 & 0.000004 \\
43     & 2459484.849830 & 0.000002 & 0.000003 \\
44     & 2459513.792354 & 0.000002 & 0.000004 \\
45     & 2459534.465589 & 0.000002 & 0.000003 \\
59     & 2459923.122341 & 0.000002 & 0.000004 \\
71     & 2460245.624785 & 0.000003 & 0.000006 \\
73     & 2460299.375195 & 0.000003 & 0.000007 \\
86     & 2460650.820263 & 0.000004 & 0.000008 \\[3pt]
\end{tabular}
\end{table}

\section*{Physical properties and distance to \targ}

\begin{table} \centering
\caption{\em Physical properties of \targ\ defined using the nominal solar units 
given by IAU 2015 Resolution B3 (ref.~\cite{Prsa+16aj}). \label{tab:absdim}}
\begin{tabular}{lr@{~$\pm$~}lr@{~$\pm$~}l}
{\em Parameter}        & \multicolumn{2}{c}{\em Star A} & \multicolumn{2}{c}{\em Star B}    \\[3pt]
Mass ratio   $M_{\rm B}/M_{\rm A}$          & \multicolumn{4}{c}{$0.9268 \pm 0.0020$}       \\
Semimajor axis of relative orbit (\Rsunnom) & \multicolumn{4}{c}{$18.416 \pm 0.020$}        \\
Mass (\Msunnom)                             &  2.5444 & 0.0086      &  2.3581 & 0.0085      \\
Radius (\Rsunnom)                           &  1.8433 & 0.0022      &  1.7658 & 0.0026      \\
Surface gravity ($\log$[cgs])               &  4.3125 & 0.0008      &  4.3169 & 0.0011      \\
Density ($\!\!$\rhosun)                     &  0.4063 & 0.0007      &  0.4285 & 0.0013      \\
Synchronous rotational velocity ($\!\!$\kms)& 22.555  & 0.027       & 21.604  & 0.032       \\
Effective temperature (K)                   & 10950   & 150         & 10350   & 150         \\
Luminosity $\log(L/\Lsunnom)$               &  1.644  & 0.024       &  1.508  & 0.025       \\
$M_{\rm bol}$ (mag)                         &  0.631  & 0.060       &  0.969  & 0.063       \\
Interstellar reddening \EBV\ (mag)          & \multicolumn{4}{c}{$0.01 \pm 0.01$}			\\
Distance (pc)                               & \multicolumn{4}{c}{$136.4 \pm 1.7$}           \\[3pt]
\end{tabular}
\end{table}


We calculated the physical properties of \targ\ using the {\sc jktabsdim} code \cite{Me++05aa} with the photometric properties from Table~\ref{tab:jktebop} and the \Porb\ from ref.\ \cite{WilsonVanhamme14mn}. We adopted $K_{\rm A} = 108.36 \pm 0.18$\kms\ and $K_{\rm B} = 116.92 \pm 0.17$\kms\ from Hubrig et al.\ \cite{Hubrig+12aa}, and the \Teff\ values from Folsom et al.\ \cite{Folsom+10mn}. The resulting physical properties are given in Table~\ref{tab:absdim}. The synchronous rotational velocities are consistent with the measured values \cite{Hubrig+06mn}.

\begin{figure}[t] \centering \includegraphics[width=\textwidth]{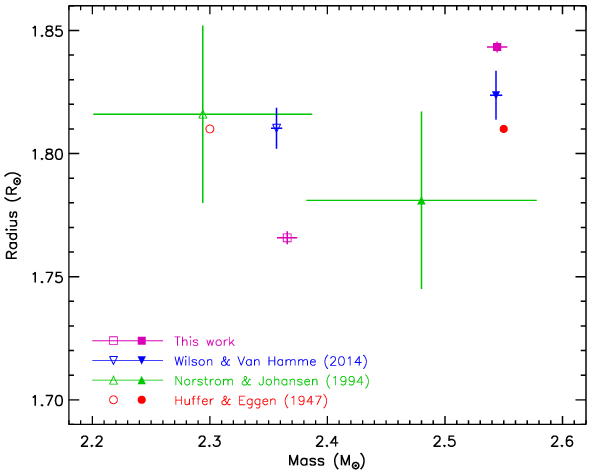} \\
\caption{\label{fig:MR} Mass--radius plot for the components of \targ\ showing 
the results from the current work and from the literature. Star~A is shown with 
filled symbols and star~B with open symbols. No uncertainties were given by 
Huffer \& Eggen \cite{HufferEggen47apj}.} \end{figure}

Fig.~\ref{fig:MR} shows measurements of the masses and radii of the components of \targ\ from this work (squares) and from the literature (triangles and circles). Our use of the new TESS data and precise velocity amplitudes from ref.~\cite{Hubrig+12aa} allows us to reach new level of precision in our measurements. Not using a SLR to constrain the ratio of the radii causes us to find a steeper mass--radius relation than previous measurements.

We determined the distance to the system using the $BV$ magnitudes from Tycho \cite{Hog+00aa}, $JHK_s$ magnitudes from 2MASS \cite{Cutri+03book} and bolometric corrections from Girardi et al.\ \cite{Girardi+02aa}. An interstellar reddening of $\EBV = 0.01 \pm 0.01$ satisfactorily equalises the distance measurements in the optical and infrared. The resulting distance to the system in the $K_s$ band is $136.4 \pm 1.7$~pc, which is 2.6$\sigma$ shorter than the \gaia\ DR3 \cite{Gaia23aa} value of $141.4 \pm 0.9$~pc.


A comparison with the theoretical predictions of the {\sc parsec} 1.2 theoretical stellar evolutionary models \cite{Bressan+12mn} finds a good agreement for a metal abundance of $Z = 0.020$ and an age of $33 \pm 3$~Myr after the zero-age main sequence. A lower $Z$ of 0.017 and an age of 59~Myr predicts a mass--radius relation steeper than observed so is disfavoured. A higher $Z$ of 0.030 can be ruled out as its zero-age main sequence predicts radii over 20$\sigma$ larger than we have measured. The \Teff\ values proposed by Takeda \cite{Takeda25raa} are higher by 200~K for star~A and 300~K for star~B, and do not match the theoretical predictions as well as the \Teff\ values we have adopted. This analysis confirms that the system contains two young main-sequence stars, and disproves earlier claims that star~B is pre-main-sequence.

\section*{Summary and conclusions}

\targ\ is a system containing a dEB of two late-B stars in an orbit of period 4.135~d, and a lower-mass outer component with a period of 23.5~yr around the inner binary. Star~A is established as a HgMn star and star~B has been found to show abundances characteristic of a weak Am star. These chemical peculiarities appear to have led to erroneous radius measurements in the past, caused by the use of a spectroscopic light ratio to help specify the ratio of the radii of the stars.

We have modelled eight sectors of data from the TESS mission using the {\sc jktebop} code, and found that the radii of the stars are very well-determined by these exceptionally good data. Combined with published spectroscopic velocity amplitudes we have determined the stars' masses to 0.35\% and their radii to 0.15\%. The properties of the system match theoretical predictions for a metallicity of $Z=0.020$ and an age of $33 \pm 3$~Myr, indicating that both components are young main-sequence stars. The distance we determine to the system is 2.6$\sigma$ shorter than the \gaia\ DR3 value; this moderate discrepancy may be due to the photospheric chemical peculiarity of the system.

We searched for pulsations by feeding the residuals of the fits to the light curves from TESS sectors 43, 44 and 45 to the {\sc period04} code \cite{LenzBreger05coast}. We found two significant frequencies, corresponding to once and twice the orbital frequency and thus explicable by slight imperfections in the light curve model. No other significant frequencies were detected up to the Nyquist limit of 359\cd. Brightness variations on the surface caused by chemical peculiarity are a plausible reason for the signals at once and twice the orbital frequency, but if so are very weak.

Our work on \targ\ therefore yields extremely precise parameter measurements which are consistent with theoretical predictions. The measured values are inconsistent with the hypothesis that star~B is a pre-main-sequence star, but do support the assertion by Takeda \cite{Takeda25raa} that spectroscopic light ratios of this system are not reliable due to the chemical peculiarity of both stars. We are left in the unusual and encouraging position of stating that no further work is needed on this system, save for perhaps a refined third-body orbit and a systematic monitoring of the photospheric abundances of the stars to search for temporal changes.


\section*{Acknowledgements}

We thank Yoichi Takeda for useful discussions.
This paper includes data collected by the TESS\ mission and obtained from the MAST data archive at the Space Telescope Science Institute (STScI). Funding for the TESS\ mission is provided by the NASA's Science Mission Directorate. STScI is operated by the Association of Universities for Research in Astronomy, Inc., under NASA contract NAS 5–26555.
This work has made use of data from the European Space Agency (ESA) mission {\it Gaia}\footnote{\texttt{https://www.cosmos.esa.int/gaia}}, processed by the {\it Gaia} Data Processing and Analysis Consortium (DPAC\footnote{\texttt{https://www.cosmos.esa.int/web/gaia/dpac/consortium}}). Funding for the DPAC has been provided by national institutions, in particular the institutions participating in the {\it Gaia} Multilateral Agreement.
The following resources were used in the course of this work: the NASA Astrophysics Data System; the SIMBAD database operated at CDS, Strasbourg, France; and the ar$\chi$iv scientific paper preprint service operated by Cornell University.



\end{document}